# Reliability updating with equality information

Daniel Straub

Engineering Risk Analysis Group, Technical University Munich (straub@tum.de)

## Abstract

In many instances, information on engineering systems can be obtained through measurements, monitoring or direct observations of system performances and can be used to update the system reliability estimate. In structural reliability analysis, such information is expressed either by inequalities (e.g. for the observation that no defect is present) or by equalities (e.g. for quantitative measurements of system characteristics). When information $Z$ is of the equality type, the a-priori probability of $Z$ is zero and most structural reliability methods (SRM) are not directly applicable to the computation of the updated reliability. Hitherto, the computation of the reliability of engineering systems conditional on equality information was performed through first- and second order approximations. In this paper, it is shown how equality information can be transformed into inequality information, which enables reliability updating by solving a standard structural system reliability problem. This approach enables the use of any SRM, including those based on simulation, for reliability updating with equality information. It is demonstrated on three numerical examples, including an application to fatigue reliability.

# 1 Introduction

In many engineering applications there is an interest in computing the probability of a rare event $E$ conditional on the observation of other events $Z_1,...,Z_n$. As an example, $E$ might represent the failure of a structural system and $Z_1,...,Z_n$ the outcomes of measurements performed on elements of that structure; an alternative example is $E$ representing the event of unacceptable deformations in a structure and $Z_1,...,Z_n$ being monitoring outcomes. Structural reliability methods (SRM) have been developed and successfully applied to this problem [1–3].

Let $\mathbf{X} = (X_1, X_2, ..., X_{n_X})$ be a set of continuous random variables, which are specified through their joint probability density function $f(\mathbf{x})$. The events $E$ and $Z_i$ are defined as domains $\Omega_E$ and $\Omega_{Z_i}$ in the outcome space of $\mathbf{X}$. The conditional probability of $E$ given $Z = \{Z_1 \cap ... \cap Z_n\}$ is then

$$\Pr(E \mid Z) = \frac{\Pr(E \cap Z)}{\Pr(Z)} = \frac{\int_{\mathbf{x} \in \{\Omega_E \cap \Omega_{Z_1} \cap ... \cap \Omega_{Z_n}\}} f(\mathbf{x}) d\mathbf{x}}{\int_{\mathbf{x} \in \{\Omega_{Z_1} \cap ... \cap \Omega_{Z_n}\}} f(\mathbf{x}) d\mathbf{x}} \qquad (1)$$

The domain $\Omega_E$ is defined in terms of continuous limit state functions $g_i(\mathbf{x})$, $i=1...m$. In the general case, it is

$$\Omega_E = \left\{ \min_{1 \leq k \leq K} \left[ \max_{i \in C_1} g_i(\mathbf{x}), ..., \max_{i \in C_K} g_i(\mathbf{x}) \right] \leq 0 \right\} \qquad (2)$$

where $C_k$ is an index set denoting the $k$-th cut set. The domain in (2) corresponds to a general system reliability problem. For $m = 1$, it becomes a component-reliability problem; for $K = 1$, it is a parallel-system reliability problem; and when each cut set contains only one index, it is a series-system reliability problem.

The domains $\Omega_{Z_i}$ are defined in terms of continuous limit state functions $h_i(\mathbf{x})$, $i = 1...n$. The information $i$ is said to be of the inequality type if it can be written as

$$\Omega_{Z_i} = \{h_i(\mathbf{x}) \leq 0\} \qquad (3)$$

and it is said to be of the equality type if it is written as

$$\Omega_{Z_i} = \{h_i(\mathbf{x}) = 0\} \qquad (4)$$

If all information is of the inequality type, then the computation of the integrals in (1) are straightforward using SRM such as first and second order reliability methods (FORM, SORM), importance sampling, directional simulation, subset simulation and others [1–4]. However, if one or more observations are of the equality type, both integrals in (1) will result in zero, because all events described by a domain of the form given in Equation (4) have zero probability. Figure 1 illustrates the problem for a component-reliability problem with two basic random variables and one observation Z. If the information is of the inequality type, the integration of $f(\mathbf{x})$ is over the areas defined by $\{h(\mathbf{x}) \leq 0\}$ and $\{g(\mathbf{x}) \leq 0 \cap h(\mathbf{x}) \leq 0\}$, which can be evaluated using SRM. If the information is of the equality type, then the domain $\Omega_Z$ becomes a surface (a line in Figure 1) in the space of $\mathbf{x}$: $\{h(\mathbf{x}) = 0\}$, and the integrals in (1) must be replaced by surface integrals, which cannot be directly evaluated using SRM.

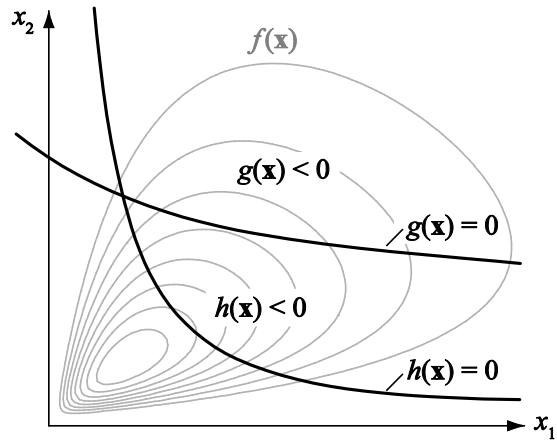

*Figure 1. Illustration of the reliability updating problem.*

Solutions to overcome the problem have been proposed in the literature. Introducing a dummy parameter $\delta$, Madsen [5] showed that

$$\Pr[g(\mathbf{X}) \leq 0 \mid h(\mathbf{X}) = 0] = \frac{\Pr[g(\mathbf{X}) \leq 0 \cap h(\mathbf{X}) = 0]}{\Pr[h(\mathbf{X}) = 0]}$$
$$= \frac{\frac{\partial}{\partial \delta}\Pr[g(\mathbf{X}) \leq 0 \cap h(\mathbf{X}) - \delta \leq 0]_{\delta=0}}{\frac{\partial}{\partial \delta}\Pr[h(\mathbf{X}) - \delta \leq 0]_{\delta=0}} \quad (5)$$

Here, the equality information $\{h(\mathbf{x})=0\}$ is replaced by the inequality event $\{h(\mathbf{x})-\delta \leq 0\}$ and the problem can be solved using SRM. The solution for the case with more than one observation is accordingly. This approach requires the partial derivatives of the probabilities with respect to the dummy parameter $\delta$. An asymptotic FORM solution exists [5], however, the error of the FORM approximation can be large. For other SRM, in particular simulation-based methods, the sensitivities in (5) must be evaluated numerically; given the numerical nature of these methods, this solution can lead to significant errors and is not practical in many cases.

Alternatively, it has been proposed to utilize first-order [6] and second-order [7] approximations to surface integration, which have been implemented in the Strurel software [8]. Such methods are efficient and, in many cases, represent a sufficiently accurate approximation. However, in cases where FORM/SORM solutions are not sufficiently accurate or in which it is difficult to identify the joint design point, these methods should not or cannot be used. Furthermore, it is often difficult to appraise the error made by the first- or second-order approximation.

In this paper, a new solution to reliability updating with equality observations is introduced. The solution is based on reformulating equality information $\{h_i(\mathbf{x})=0\}$ into a likelihood function, which can then be expressed as equivalent inequality information $\{h_{e,i}(\mathbf{x}_+) \leq 0\}$ in the outcome space of an altered set of variables $\mathbf{X}_+$. It is suitable for all SRM, including those based on simulation, such as importance sampling and directional simulation, which makes it powerful for problems where FORM /SORM approximations are inaccurate or difficult to obtain. The proposed approach is demonstrated and verified on numerical examples, including an application to reliability updating of a structural element subject to fatigue crack growth.

## 2  Likelihood function

In Bayesian analysis, the effect of information $Z$ (observations, data) on the uncertain parameters $\Theta$ is expressed through the likelihood function

$$L(\theta) = \Pr(Z \mid \Theta = \theta) \tag{6}$$

For most types of information, it is straightforward to directly formulate the likelihood function without defining the corresponding limit state function first. Modelers used to describing observations by means of limit state functions $h(\mathbf{x})$ can always derive the likelihood function from $h(\mathbf{x})$, as described in the following paragraph.

First, we note that by interpreting the basic random variables $\mathbf{X}$ in Eqs. (1) to (4) as the model parameters $\Theta$ in Eq. (6), the likelihood function corresponding to an equality event $\{h(\mathbf{x}) = 0\}$ becomes the Dirac delta function with argument $h(\mathbf{x})$. Such a representation is not useful for computation with SRM. An alternative way of establishing a likelihood function equivalent to $\{h(\mathbf{x}) = 0\}$ proceeds by partitioning the random variables $\mathbf{X}$ into variables $\mathbf{X}_g$ and variables $\mathbf{X}_h$, whereby the latter contain some or all of the variables in $\mathbf{X}$ that appear exclusively in $h(\mathbf{x})$ and the former contain the remaining variables in $\mathbf{X}$. A typical example for an element of $\mathbf{X}_h$ is a random variable describing measurement uncertainty. If $X_h$ is a scalar variable, it is possible to describe the likelihood function of $\mathbf{X}_g$ given the information $Z$ as

$$L(\mathbf{x}_g) \propto \Pr(Z | \mathbf{X}_g = \mathbf{x}_g)$$
$$\propto \sum_{j=1}^{n_h} f_{X_h}[\hat{x}_{h,j}(\mathbf{x}_g)] \quad (7)$$

where $f_{X_h}$ is probability density function of $X_h$ and $\hat{x}_{h,j}(\mathbf{x}_g)$ are the $n_h$ roots of the equation $h(x_h, \mathbf{x}_g) = 0$ for given $\mathbf{x}_g$. When $\mathbf{X}_h$ is a vector, the summation in Eq. (7) must be replaced by an integration over the surface $h(\mathbf{x}_h, \mathbf{x}_g) = 0$. Eq. (7) is valid only if the $\mathbf{X}_h$ are statistically independent of $\mathbf{X}_g$, which is commonly the case for variables $\mathbf{X}_h$ that represent measurement and observation uncertainty; otherwise the PDF $f_{\mathbf{X}_h}$ in (7) should be replaced with the conditional PDF of $X_h$ given $\mathbf{X}_g$, $f_{X_h|\mathbf{X}_g}$.

As stated earlier, for most type of equality information, the likelihood function according to Eq. (7) can be directly established without defining $h(\mathbf{x})$ first. In particular, if $Z$ describes the measurement $s_m$ of a system characteristic $s(\mathbf{x})$ with an additive measurement error $\varepsilon$, then the likelihood function is $L(\mathbf{x}_g) = f_\varepsilon[s_m - s(\mathbf{x}_g)]$, with $f_\varepsilon$ being the PDF of the measurement error and $\mathbf{X}_g$ containing all variables in $\mathbf{X}$ except $\varepsilon$. [The corresponding equality limit state function would be $h(\mathbf{x}_g, \varepsilon) = s(\mathbf{x}_g) - s_m + \varepsilon$].

It is possible that a limit state function describing equality information does not contain a basic random variable that is not also part of the limit state functions $g_i(\mathbf{x})$, $i=1…m$. In this case, an approximate likelihood function $L(\mathbf{x}_g)$ is obtained by adding an additive measurement error with zero mean and a standard deviation that is small compared to the uncertainty in the observed quantities.

When several events $Z_1,...,Z_n$ are observed, it is alternatively possible to establish separate likelihood functions $L_1,…,L_n$ or a single likelihood function $L$ describing the combined observations. For the common case that each observation is subject to an independent measurement error $\varepsilon_i$, it is $L(\mathbf{x}_g) = \prod_{i=1}^{n} L_i(\mathbf{x}_g)$, where $\mathbf{X}_g$ contains all basic random variables of the problem excluding the $\varepsilon_i$, $i = 1,…,n$.

## 3 Reliability updating with equality information expressed as likelihood functions

In this section, it will be shown how equality information $Z$ expressed by a likelihood function $L(\mathbf{x}_g)$ of the form given in Eq. (7) can be described as inequality information. ($\mathbf{X}_g$ includes all basic random variables that appear in the limit state functions $g_i(\mathbf{x})$, $i=1…m$ as introduced in the previous section.) The formulations presented hereafter are equally valid for the case of several information events $Z_1,...,Z_n$ if these are described by a common likelihood function $L(\mathbf{x}_g)$.

It is noted that the following identity holds for any likelihood function $L(\mathbf{x}_g)$:

$$L(\mathbf{x}_g) = \frac{1}{c}\Pr\{U - \Phi^{-1}[cL(\mathbf{x}_g)] \leq 0\} \tag{8}$$

where $U$ is a standard Normal random variable, $\Phi^{-1}$ is the inverse standard Normal cumulative distribution function and $c$ is a positive constant that is chosen to ensure that $0 \leq cL(\mathbf{x}_g) \leq 1$ for all $\mathbf{x}_g$. To verify Eq. (8) it suffices to realize that $\Pr\{U - \Phi^{-1}[p] \leq 0\} = \Phi\{\Phi^{-1}[p]\} = p$ with $0 \leq p \leq 1$. The formulation in Eq. (8) is inspired by the nested reliability formulation originally proposed in [9]. The relation (8) enables expressing the likelihood function by the limit state function

$$h_e(\mathbf{x}_g, u) = u - \Phi^{-1}[cL(\mathbf{x}_g)] \tag{9}$$

and the corresponding (inequality) domain

$$\Omega_{Ze} = \{h_e(\mathbf{x}_g, u) \leq 0\} \tag{10}$$

With these definitions, Eq. (8) can be further developed into

$$L(\mathbf{x}_g) = \frac{1}{c} \int_{\mathbf{x}_g, u \in \Omega_{Ze}} \varphi(u) du \tag{11}$$

where $\varphi(\cdot)$ is the standard Normal PDF. Introducing $\alpha$ as the proportionality constant in Eq. (7), we can write

$$\Pr(Z | \mathbf{X}_g = \mathbf{x}_g) = \frac{\alpha}{c} \int_{\mathbf{x}_g, u \in \Omega_{Ze}} \varphi(u) du \tag{12}$$

It follows that the probability of the observed event $Z$ is

$$\begin{aligned} \Pr(Z) &= \int_{\mathbf{X}_g} \Pr(Z | \mathbf{X}_g = \mathbf{x}_g) f(\mathbf{x}_g) d\mathbf{x}_g \\ &= \frac{\alpha}{c} \int_{\mathbf{x}_g, u \in \Omega_{Ze}} \varphi(u) f(\mathbf{x}_g) du d\mathbf{x}_g \end{aligned} \tag{13}$$

Similarly, we have that

$$\begin{aligned} \Pr(E \cap Z) &= \int_{\mathbf{X}_g} \Pr(E | \mathbf{X}_g = \mathbf{x}_g) \Pr(Z | \mathbf{X}_g = \mathbf{x}_g) f(\mathbf{x}_g) d\mathbf{x}_g \\ &= \frac{\alpha}{c} \int_{\mathbf{x}_g, u \in \{\Omega_E \cap \Omega_{Ze}\}} \varphi(u) f(\mathbf{x}_g) du d\mathbf{x}_g \end{aligned} \tag{14}$$

The conditional probability of $E$ given $Z$ is thus

$$\Pr(E | Z) = \frac{\Pr(E \cap Z)}{\Pr(Z)} = \frac{\int_{\mathbf{x}_g, u \in \{\Omega_E \cap \Omega_{Ze}\}} \varphi(u) f(\mathbf{x}_g) du d\mathbf{x}_g}{\int_{\mathbf{x}_g, u \in \Omega_{Ze}} \varphi(u) f(\mathbf{x}_g) du d\mathbf{x}_g} \tag{15}$$

Note that the proportionality constant $\alpha$ vanishes in (15).

For notational convenience, define $\mathbf{X}_+ = [\mathbf{X}_g; U]$. This vector represents the augmented space of basic random variables. Eq. (15) is thus rewritten to

$$\Pr(E \mid Z) = \frac{\int_{\mathbf{x}_+ \in \{\Omega_E \cap \Omega_{Ze}\}} f(\mathbf{x}_+) d\mathbf{x}_+}{\int_{\mathbf{x}_+ \in \Omega_{Ze}} f(\mathbf{x}_+) d\mathbf{x}_+} \qquad (16)$$

The relation in Eq. (16) is easily extended to the case of multiple equality information $Z_1, \ldots, Z_n$ described by separate likelihood functions $L_1, \ldots, L_n$. For each likelihood function $L_i$, the corresponding inequality domain $\Omega_{Ze_i}(\mathbf{x}_+)$ is defined according to Eq. (10), with a separate random variable $U_i$ for each $i = 1, \ldots, n$. Therefore, the augmented space now consists of $\mathbf{X}_+ = [\mathbf{X}_g; U_1; \ldots; U_n]$. The conditional probability of $E$ is

$$\begin{aligned}\Pr(E \mid Z_1, \ldots, Z_n) &= \frac{\Pr(E \cap Z_1 \cap \ldots \cap Z_n)}{\Pr(Z_1 \cap \ldots \cap Z_n)} \\ &= \frac{\int_{\mathbf{x}_+ \in \{\Omega_E \cap \Omega_{Ze_1} \cap \ldots \cap \Omega_{Ze_n}\}} f(\mathbf{x}_+) d\mathbf{x}_+}{\int_{\mathbf{x}_+ \in \{\Omega_{Ze_1} \cap \ldots \cap \Omega_{Ze_n}\}} f(\mathbf{x}_+) d\mathbf{x}_+}\end{aligned} \qquad (17)$$

A main advantage of the proposed approach is that the integrals in Eqs. (16) and (17) can be solved using any SRM, including, in particular, simulation methods. If the associated probabilities are large and the evaluation of the limit state function is computationally inexpensive, crude Monte Carlo simulation can be applied, which, in contrast to other methods, provides unbiased results together with an estimate of the associated error; otherwise the use of advanced simulation methods, such as importance sampling or subset simulation, is recommended. FORM/SORM can be applied, but requires due attention to the shape of the limit state functions around the joint design points.

## 4  Numerical examples

The first two numerical examples are evaluated with an importance sampling method that uses line searches from the hyperplane perpendicular to the joint design point vector at the origin in standard Normal space. The method was originally proposed in [10] and is summarized in [2], where it is referred to as Axis-Parallel Importance Sampling (APIS). With $d(v_i)$ being the distance between the $i^{\text{th}}$ sample on the hyperplane (with location $v_i$) and its corresponding projection on the limit state surface, the probability estimate is obtained from $n_{ls}$ line searches by

$$p \approx \frac{1}{n_{ls}} \sum_{i=1}^{n_{ls}} \Phi[-d(v_i)] \frac{\varphi(v_i)}{\psi(v_i)} \tag{18}$$

where $\psi(v_i)$ is the sampling density.

All results are expressed in terms of the reliability index $\beta$, which is related to the probability estimate $p$ by $\beta = -\Phi^{-1}(p)$.

## 4.1 Low-dimensional example

This basic example facilitates graphical illustration of the approach, since it involves only two random variables. The event of interest $E$ is defined by the domain $\Omega_E(\mathbf{x}) = \{g(r) \leq 0\}$ through the limit state function

$$g(r) = r - s \tag{19}$$

The loading $s$ is deterministic and takes value $s = 2.0$. The capacity $R$ has cumulative distribution function

$$F_R(r) = 1 - \exp\left[-\left(\frac{r}{v}\right)^k\right] \tag{20}$$

This is the Weibull distribution with shape parameter $k = 3.0$ and scale parameter $v = 10$. The unconditional probability of failure is $F_R(2.0) = 0.008$, and the corresponding reliability index is $\beta = 2.41$.

A measurement of the capacity is performed, and results in $r_m = 6.0$. The measurement has additive standard Normal distributed error $\varepsilon_m$. The limit state function describing this equality information is $h(r, \varepsilon_m) = r - r_m + \varepsilon_m$ and the corresponding likelihood function is

$$L(r) = \varphi(r_m - r) \tag{21}$$

By inserting Eq. (21) into Eq. (9), the equivalent inequality limit state function is obtained as

$$h_e(r, u) = u - \Phi^{-1}[c\varphi(r_m - r)] \tag{22}$$

Since $\varphi(r_m - r) < 1$, we can set the constant as $c = 1$. The limit state surfaces $h_e(r,u) = 0$ and $g(r) = 0$ are shown in Figure 2.

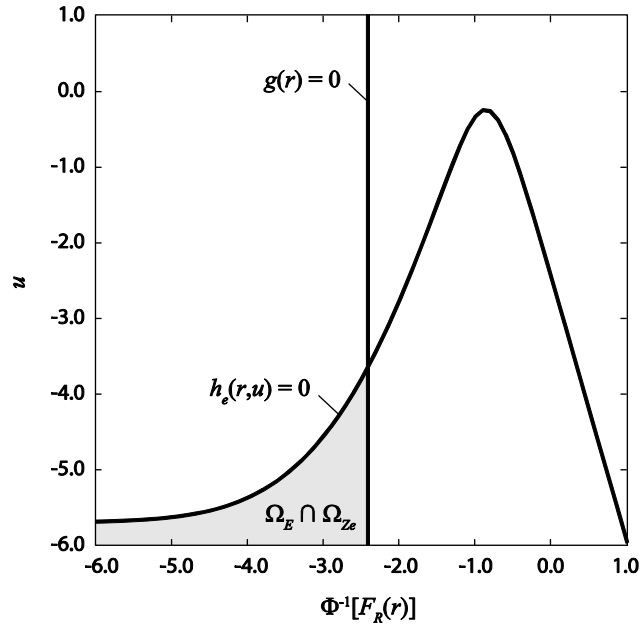

*Figure 2. Limit state surfaces of the problem with the equivalent inequality information, in standard Normal space.*

The conditional reliability index given the measurement obtained with the proposed approach utilizing the limit state function (22) and evaluated with the APIS method with 500 line searches is [4.47,4.53] (the range is obtained from repeating the simulations 10 times with different seed values for the random number generator). This compares well with the exact solution $\beta = 4.49$ obtained by numerical integration from $\Pr(F|Z) = \int_0^s f_R(r)L(r)dr \Big/ \int_0^\infty f_R(r)L(r)dr$.

It is observed from Figure 2 that the limit state surface $h_e(r,u) = 0$ is strongly non-linear. This is commonly the case for the equivalent inequality limit state functions. Therefore, FORM/SORM algorithms should be applied carefully in conjunction with the proposed approach, in particular for problems in higher dimensions. For the present example, with limit state function (22), the FORM solution is $\beta_{FORM} = 4.69$ and the SORM solution is $\beta_{SORM} = 4.60$. As evident from Figure 2, the (small) error occurs mainly in the computation of the denominator in Eq. (16).

## 4.2 Linear and Normal case

Here, linear limit state functions with exclusively Normal-distributed random variables are considered, for which an analytical solution exists. The limit state function for the event of interest is

$$g(\mathbf{x}) = \mathbf{a}\mathbf{x} \tag{23}$$

where $\mathbf{a} = [2,3,6,4,-1,-2,-4,-4]$ are the parameters and $\mathbf{X} = [X_1,...,X_8]^T$ are statistically independent Normal random variables with means $\mu_X = 10$ and standard deviation $\sigma_X = 2$. Equality information described by the following limit state functions is considered:

$$h_1(x_1, x_2, \varepsilon_1) = x_1 + x_2 - 20 + \varepsilon_1$$

$$h_2(x_2, x_3, \varepsilon_2) = x_2 + x_3 - 20 + \varepsilon_2 \tag{24}$$

$$h_3(x_3, x_4, \varepsilon_3) = x_3 + x_4 - 20 + \varepsilon_3$$

wherein the $\varepsilon_i, i = 1,2,3$ are statistically independent standard Normal random variables. The corresponding likelihood functions are

$$L_i(x_i, x_{i+1}) = \varphi(20 - x_i - x_{i+1}), \quad i = 1,2,3 \tag{25}$$

By inserting into Eq. (9), the equivalent inequality limit state functions are obtained as

$$h_{e_i}(\mathbf{x}_+) = u_i - \Phi[c_i \varphi(20 - x_i - x_{i+1})], \quad i = 1,2,3 \tag{26}$$

Since $\varphi(20 - x_i - x_{i+1}) < 1$, the constants are set as $c_i = 1$.

The conditional reliability index is calculated as $\beta = [3.02, 3.08]$ using APIS with 500 line searches. (The associated number of limit state function calls is approx. $10^4$.) For comparison, the analytical solution for the updated reliability index is obtained as $\beta = 3.07$. The FORM and SORM solutions with the equivalent inequality limit state functions (26) are $\beta_{FORM} = 3.51$ and $\beta_{SORM} = 2.95$, reflecting the significant non-linearity of the limit state surfaces.

## 4.3 Fatigue crack growth with measurements

This example illustrates the application of the proposed approach in updating deterioration reliability. It demonstrates how the approach can be combined with crude Monte Carlo simulation (MCS). For illustration purposes, a crack growth model from [1] is considered. The rate of crack growth is described by Paris' law as

$$\frac{d a(n)}{dn} = C\left[\Delta S \sqrt{\pi a(n)}\right]^m \tag{27}$$

$a$ is the crack length, $n$ is the number of stress cycles, $\Delta S$ is the stress range per cycle (constant stress amplitudes are assumed) and $C$ and $m$ are empirically determined model parameters. In this formulation of Paris' law, the geometry correction factor is one, which in theory corresponds to the case of a crack in a plate with infinite size. With the boundary condition $a(n=0) = a_0$, this differential equation can be solved for the crack size as a function of the number of cycles $n$, [1]:

$$a(n) = \left[\left(1 - \frac{m}{2}\right) C \Delta S^m \pi^{m/2} n + a_0^{(1-m/2)}\right]^{(1-m/2)^{-1}} \tag{28}$$

The event of failure is described by the limit state function $g$ as a function of $a(n)$ and the critical crack length $a_c$:

$$g = a_c - a(n) \tag{29}$$

The model parameters are summarized in Table 1.

*Table 1. Parameters of the crack growth example.*

| Variable | Distribution | Mean | St. deviation | Correlation |
| --- | --- | --- | --- | --- |
| $a_0$ [mm] | Exponential | 1 | 1 | - |
| $a_c$ [mm] | Deterministic | 50 | - | - |
| $\Delta S$ [Nmm$^{-2}$] | Normal | 60 | 10 | - |
| $\ln(C), m$ [*] | Bi-Normal | [-33; 3.5] | [0.47; 0.3] | $\rho_{\ln(C),m} = -0.9$ |

\* dimension corresponding to N and mm

Two measurements of crack depth are considered. The corresponding limit state functions representing the equality information are

$$h_i(\mathbf{x}) = a(\mathbf{x}, n_i) - a_{m,i} + \varepsilon_i, \quad i = 1,2 \qquad (30)$$

wherein $n_i$ are the number of stress cycles up to the $i$th measurement, $a_{m,i}$ are the crack depths measurements and $\varepsilon_i$ are the measurement errors, modeled as statistically independent standard Normal random variables. The corresponding equivalent inequality limit state functions are

$$h_{e_i}(\mathbf{x}_+) = u_i - \Phi\{c_i \, \varphi[a_{m,i} - a(\mathbf{x}, n_i)]\}, \quad i = 1,2 \qquad (31)$$

Here, the constants are selected as $c_i = 1$. The measured crack depths are

$a_{m,i} = 0.5\text{mm}$ at $n_i = 3 \cdot 10^5$ stress cycles

$a_{m,i} = 3.0\text{mm}$ at $n_i = 2 \cdot 10^6$ stress cycles

For numerical evaluation, MCS with $10^6$ samples is utilized. For comparison, additionally the dynamic Bayesian networks (DBN) framework for stochastic deterioration modeling suggested in [11] is applied, as well as a second-order approximation to the surface integral, which is based on the methods described in [7] and utilizes the equality limit state functions (30). The results are presented in Figure 3.

It is pointed out that the results obtained with MCS in combination with the proposed equivalent inequality limit state function do not include any approximation and are unbiased. The 95% confidence interval of the MCS, which indicates the sampling error, is small, as shown in Figure 3. A smaller number of samples would be sufficient for most practical purposes.

The results obtained with the DBN model match closely those of the MCS, with a small deviation after the second measurement, where the approximations made in the DBN model appear to have a stronger effect. The second-order results strongly underestimate the true reliability index. For $n \geq 4 \cdot 10^6$, no results could be obtained with this approach, due to algorithmic difficulties in the design point search (this is not a fundamental problem and could be overcome by improving the applied optimization algorithm, but it is quite common in practical implementations of the second-order approach).

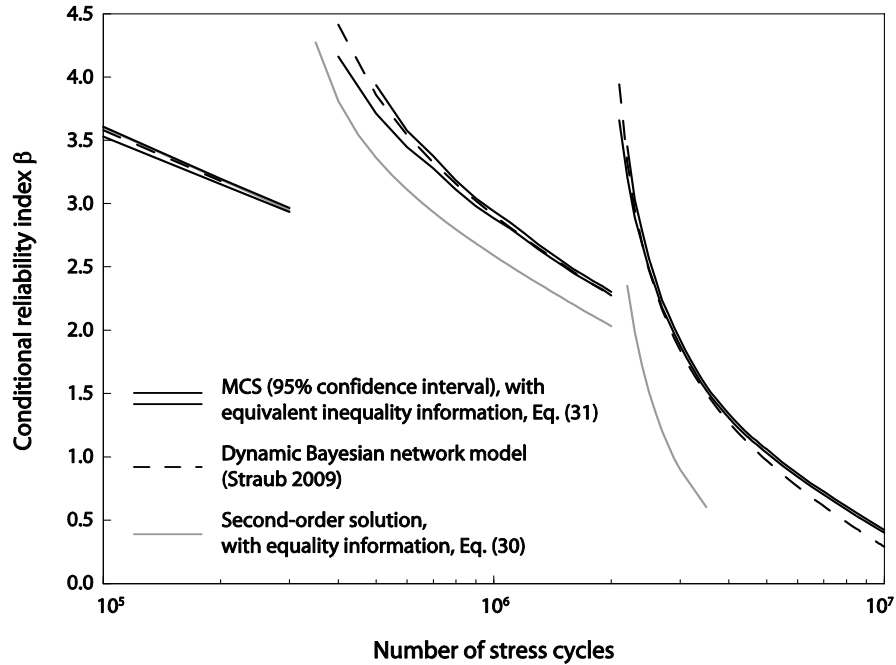

*Figure 3. Reliability index after n stress cycles, conditional on the measurement results up to n cycles.*

## 5 Concluding remarks

Equality information is described through equalities in the space of the (continuous) basic random variables of a reliability problem, i.e., it encompasses all information that has zero-probability a-priori. Typical examples of such information are measurements, monitoring outcomes or observations of system performances. Reliability updating with such type of information is a relevant problem in many fields of structural reliability and risk analysis.

In this paper, it is shown that equality information can be transformed into inequality information, i.e., information that has non-zero probability of occurrence a-priori. The approach proceeds by formulating likelihood functions for the equality information and then describing them by a domain in the combined space of the basic random variables and auxiliary random variables. The benefit of this transformation lies in the fact that equivalent inequality information is amenable to simulation-based structural reliability methods, which hitherto were difficult or impossible to apply for reliability updating with equality information.

No approximations are made in the transformation from the equality to the equivalent inequality information. If evaluated with Monte Carlo simulation, the approach, unlike existing methods, allows providing an unbiased solution of the updated reliability, together with an error estimate, as demonstrated in example 3. However, in many applications the probabilities of the events of interest are small, in which case the use of advanced simulation methods, including importance sampling or subset simulation, is recommended. The application of FORM/SORM to evaluate the updated reliability with the equivalent inequality information according to Eq. (17) is advocated only if the shape of the limit state surfaces around the joint design points can be determined. Because limit state surfaces representing the equivalent inequality information are considerably non-linear, as illustrated in Figure 2, FORM/SORM approximations may provide inaccurate results.

In conclusion, the proposed approach presents a novel solution to reliability updating when measurements, monitoring and other information of the equality type are available. Its key advantage over existing methods is that it enables the use of simulation-based methods for the evaluation of the updated reliability. The implementation of the approach is straightforward, since the only additional step required, as compared with classical structural reliability analysis, is the formulation of the equivalent inequality limit state functions.